\documentclass{aa}  

\usepackage{graphicx}
\usepackage{txfonts}
\usepackage{hyperref}
\usepackage{amsmath}
\usepackage{units}
\usepackage{footnote}
\usepackage{xcolor}
\usepackage{multirow}

\begin{document}

   \title{Assessing the habitability of planets with Earth-like atmospheres with 1D and 3D climate modeling}

   \author{M. Godolt \inst{1,2} \and
J. L. Grenfell \inst{2} \and 
D. Kitzmann \inst{3} \and 
M. Kunze \inst{4} \and 
U. Langematz \inst{4} \and 
A. B. C. Patzer\inst{1} \and 
H. Rauer \inst{1,2} \and 
B. Stracke \inst{2}
          }

   \institute{Centre of Astronomy and Astrophysics, Technische Universit\"at Berlin, Hardenbergstr. 36, 10623 Berlin, Germany\\
   \email{godolt@astro.physik.tu-berlin.de}
   \and Institute of Planetary Research, German Aerospace Center (DLR), Rutherfordstr.~2, 12489 Berlin, Germany
\and
Center for Space and Habitability (CSH), Sidlerstrasse 5, 3012 Bern, Switzerland
\and
Institute of Meteorology, Freie Universit\"at Berlin, Carl-Heinrich-Becker-Weg 6-10, 12165 Berlin, Germany
             }

  \abstract
   {The habitable zone (HZ) describes the range of orbital distances around a star where the existence of liquid water on the surface of an Earth-like planet is in principle possible. The applicability of one-dimensional (1D) climate models for the estimation of the HZ boundaries has been questioned by recent three-dimensional (3D) climate studies. While 3D studies can calculate the water vapor, ice albedo, and cloud feedback self-consistently and therefore allow for a deeper understanding and the identification of relevant climate processes, 1D model studies rely on fewer model assumptions and can be more easily applied to the large parameter space possible for extrasolar planets.
}
   {
We evaluate the applicability of 1D climate models to estimate the potential habitability of Earth-like extrasolar planets by comparing our 1D model results to those of  3D climate studies in the literature. We vary the two important planetary properties, surface albedo and relative humidity, in the 1D model. These depend on climate feedbacks that are not treated self-consistently in most 1D models. 
}
   {We applied a cloud-free 1D radiative-convective climate model to calculate the climate of Earth-like planets around different types of main-sequence stars with varying surface albedo and relative humidity profile. We compared the results to those of 3D model calculations available in the literature and investigated to what extent the 1D model can approximate the surface temperatures calculated by the 3D models.
}
   {The 1D parameter study results in a large range of climates possible for an Earth-sized planet with an Earth-like atmosphere and water reservoir at a certain stellar insolation. At some stellar insolations the full spectrum of climate states could be realized, i.e.,~uninhabitable conditions due to surface temperatures that are too high or too low as well as habitable surface conditions, depending only on the relative humidity and surface albedo assumed. 
When treating the surface albedo and the relative humidity profile as parameters in 1D model studies and using the habitability constraints found by recent 3D modeling studies, the same conclusions about the potential habitability of a planet can be drawn as from 3D model calculations.}
   {}

   \keywords{astrobiology, planets and satellites: atmospheres, terrestial planets }

   \maketitle
%

\section{Introduction}
The classical habitable zone (HZ), as defined, for example, by \cite{kasting1993hz}, is a concept that permits the determination of the range of orbital distances for which an Earth-like planet around different types of main-sequence stars may in principle have liquid water on its surface over an extended period of time. It is a valuable concept for determining where one should look for potential life on extrasolar planets, since life as we know it needs liquid water at least for a part of its life cycle and surface life is likely to be more easily detectable, as it may alter the planetary atmosphere. The boundaries of the HZ determined by \cite{kasting1993hz} have recently been revised by including, for example, updates in the radiative transfer \citep{Kopparapu2013}, atmospheric dynamics \citep{Leconte2013b,Wolf2013,Wolf2015}, planets with smaller water reservoirs \citep{Abe2011,Leconte2013, Zsom2013}, or the influence of different rotation rates \citep{Yang2013,Yang2014} as well as different atmospheric 
compositions \citep[e.g.,][]{Pierrehumbert2011}.
The large spread in orbital distances for the inner edge of the HZ resulting from these studies led to some discussions of which boundaries should be taken into account when planning future missions or instruments aimed at the study of such objects or for the determination of the occurrence rate of Earth-like planets \citep[see, e.g.,][]{Kasting2014}.  The inner boundaries of the HZ from the 1D modeling study by \cite{Kopparapu2013} are the most conservative in the sense that present Earth is close to the edge of this HZ. Such a strict HZ boundary definition is valuable, since it provides, for example, lower limits on the number of potentially habitable planets that could be observed with a given mission design and is, therefore, important to facilitate the achievement of a certain mission goal. The modeling studies by \cite{Leconte2013b}, \cite{Wolf2013}, \cite{Wolf2015}, and \cite{Yang2014} have tried to assess the inner HZ boundary with 3D models, treating the water vapor feedback self-consistently. They 
showed that the inner edge of the HZ is located closer to the star, when accounting for the water vapor feedback cycle, since relative humidities of 100\% are not reached when including atmospheric dynamics. Additionally, a new stabilizing feedback by clouds was found by \cite{Yang2013,Yang2014} for slowly rotating planets, which would have been impossible to identify by 1D modeling studies.

While 1D model calculations can be simply applied for a wide range of stellar and planetary parameters, the physical processes leading to a certain climate state can be better understood by applying a 3D model since climate is clearly a 3D phenomenon. 
However, 3D climate calculations require a huge set of boundary conditions, such as continental distribution, orography, obliquity, rotation rate, and oceanic heat transport. These are, in addition to the more obvious parameters such as atmospheric mass and composition, parameters that we do not know for potentially habitable rocky extrasolar planets. While it is important to understand how these parameters may influence the climate, such as studied~for the rotation rate  by \cite{Yang2014,Yang2013} and \cite{DelGenio1987} or the oceanic heat transport \citep{Yang2013,Cullum2014}, accounting for all possible combinations of boundary parameters in addition to the unknown atmospheric composition and mass is rather impractical. Hence, 1D models are important to assess the main aspects of planetary climate, or in our case habitability, but also for atmospheric biomarker studies with coupled climate chemistry models. Therefore, exoplanet atmosphere science would benefit if the findings of recent 3D model 
studies could be considered in 1D model calculations.\\ 

So far, 1D modeling studies of the HZ boundaries by \cite{kasting1993hz} and \cite{Kopparapu2013} have utilized a fixed relative humidity of 100\% and a surface albedo tuned to reproduce the mean temperature of the Earth when utilizing a measured relative humidity profile of Earth. The assumption of a fully saturated atmosphere maximizes the greenhouse effect by water vapor \citep[as also mentioned by ][]{kasting1993hz} and, thereby, overestimates the water vapor feedback as atmospheric circulation leads to dehydration as also shown by 3D modeling studies \citep{Leconte2013b}. Such model assumptions, nevertheless, yield  conservative inner edge distances to the star where a planet with an Earth-like water reservoir may be habitable. \\

\cite{Zsom2013} studied  the inner edge of the habitable zone with a 1D model for planets with rather low relative humidity and different surface albedos,
 showing that the inner edge may be much closer to the star for dry planets; this was also shown by 3D modeling studies \citep{Abe2011,Leconte2013}. The minimal distances to the star derived by these studies are, however, different. In comparison, \cite{Abe2011} obtain a rather large orbital distance needed by a dry planet to stay habitable with a stellar insolation of 1.7 times the solar value, while \cite{Zsom2013} obtain a very small orbital distance with a stellar insolation about 7 times solar. For their smallest distance, however, \cite{Zsom2013} assume a high surface albedo of 0.8, which is, as the authors state, rather improbable for a dry land planet. When comparing the results by \cite{Zsom2013} with a surface albedo of 0.2 to those of \cite{Leconte2013} for GJ581c, the discrepancy between the 1D and 3D model results is smaller, yielding habitable surface conditions for stellar insolations up to about 2.9 and 2.5 times the solar value, respectively. \\

Most 1D atmosphere calculations assume a distribution of the water vapor in the atmosphere via relative humidity parameterizations \citep[e.g.,][]{Rugheimer2015,vonParis2015,Grenfell2014, Zsom2013,Wordsworth2013, Segura2003}, rather than compute the hydrological cycle self-consistently. In their 1D modeling study, \cite{Popp2015} calculate the water vapor distribution self-consistently without using an assumption about the relative humidity profile. These authors utilized the column setup of the ECHAM6 3D climate model \citep{ECHAM6}, showing that high relative humidities up to 100\% result when neglecting the horizontal water vapor transport by atmospheric dynamics.\\

The relative humidity, as well as the surface albedo, strongly impact the climate of a planet. They are directly linked with the water vapor and ice albedo climate feedbacks. While the qualitative impact of these climate feedbacks is mostly understood, determining their quantitative impact on climate is not an easy task. This depends on local surface properties, atmospheric and oceanic dynamics, and therefore also on the atmospheric composition and stellar spectral flux distribution. However, one may argue that surface albedo and relative humidity can vary only over a limited range, when focusing on planets with a sufficient liquid water reservoir similar to the Earth, which is in phase equilibrium with the atmosphere.\\

Three-dimensional modeling studies by \cite{Shields2014}, \cite{Boschi2013}, and \cite{Godolt2015} have shown that for a certain stellar insolation very different climate states, both habitable and uninhabitable, are possible, depending~on the initial climate state or the assumptions of the oceanic heat transport, for example. For these different climate states, the water vapor distribution and surface albedo are very different. This possibility of climate bistabilities is usually not considered when estimating the width of the habitable zone, since it only focuses on determining the HZ boundaries and not on determining the particular climate state of a certain terrestrial planet. When discussing the habitability of a rocky planet within the habitable zone, however, these bistabilities should be kept in mind.\\

In this study, we explore the range of possible habitable climates of Earth-like planets around different types of central stars with a cloud-free 1D climate model. The range of different climates is found by varying, firstly, the relative humidity profile between one measured for Earth, \cite{MW1967} and a fully saturated atmosphere. Secondly, the surface albedo of the planet is varied to simulate planets completely covered by an ocean, which have a low albedo when they are ice free and have a high albedo when the surface water is completely frozen. 
The results of the 1D model calculations are compared to those of 3D model studies from the literature to estimate to what extent the 3D model results can be assessed by the 1D model.\\

The paper is structured as follows: The next section gives details about the 1D climate model used  (\ref{sec:model}) and the various model calculations conducted  (\ref{sec:runs}). In the results section we discuss the results of the 1D model for different relative humidities and surface albedos (\ref{sec:RH_alb}), and show how they compare with 3D model results from the literature. We show a more detailed comparison with the 3D model results from \cite{Kunze2014} and \cite{Godolt2015} in Sec.~\ref{sec:3d_comp}, followed by  a discussion on the results and on future work required. The paper closes with a short summary.


\section{Method}
\subsection{Model description}
\label{sec:model}
 We apply a cloud-free 1D radiative-convective climate model, originally based on \cite{Kasting1984}\footnote{as provided on the website of the Virtual Planetary Laboratory: http://vpl.astro.washington.edu/sci/AntiModels/models09.html}, to assess for which stellar insolations Earth-like planets can be habitable
under the assumption of different surface albedos and relative humidity profiles. Improvements to the model have been implemented by various authors and are described by \cite{vonParis2008, vonParis2010, vonParis2015} and references therein. 
The model calculates the steady-state, vertical, global mean atmospheric temperature and water vapor profiles from the surface up to a height corresponding to a pressure of \unit[6.6$\cdot$10$^{-5}$]{bar}. The temperature profile is calculated from energy transport by radiative transfer and convective adjustment. The calculation of the radiative transfer through the atmosphere is split into a shortwave and a longwave wavelength regime. In the shortwave wavelength regime, from 237 nm to \unit[4.5]{$\mu$m}, the radiative transfer equation is solved in 38 spectral bands, using a $\delta$-Eddington 2-stream Eddington approximation \citep{Toon1989} and correlated-k exponential sums treating the absorption and scattering of the incident stellar light by the planetary atmosphere. Absorption by water vapor (H$_2$O), carbon dioxide (CO$_2$), molecular oxygen (O$_2$), ozone (O$_3$), methane (CH$_4$), and Rayleigh scattering by molecular nitrogen (N$_2$) and O$_2$, H$_2$O, CO$_2$, and CH$_4$ are considered.  In the 
longwave wavelength regime, ranging from 1 to \unit[500]{$\mu$m}, the radiative transfer equation is solved in 25 bands accounting for the thermal emission by the planetary surface and atmosphere and its absorption using the correlated-k approach. In the longwave wavelength regime absorption by H$_2$O, CO$_2$, O$_3$, and CH$_4$ are considered. The radiative transfer as described in \cite{vonParis2015} is applied.
Wherever the radiative lapse rate is larger than the adiabatic lapse rate, convective adjustment is performed to dry or moist adiabatic conditions ranging down to the surface. The water vapor profile ($C_{\mathrm{H_2O}}(p)$) in the troposphere is calculated from the temperature profile ($T(p)$) and a relative humidity ($RH$) parametrization by
\begin{equation}
C_{\mathrm{H_2O}}(p)=\frac{p_{\mathrm{sat,H_2O}}(T(p))}{p}RH(p) \text{,}
\end{equation}
where $p_{\mathrm{sat,H_2O}}$ is the saturation vapor pressure of water and $p$ the height dependent atmospheric pressure. 
Either a relative humidity profile as measured for the modern Earth by \cite{MW1967} or a relative humidity of 100\% is applied,
\begin{equation}
RH(p)=\begin{cases} RH_{\mathrm{surf}}\frac{\frac{p}{p_{\mathrm{surf}}}-0.02}{0.98}  \rightarrow \text{RH MW} \\ 100\% \hspace{1.35cm}\rightarrow \text{RH 100}\\ \end{cases} \text{,}
\end{equation}
where $RH_{\mathrm{surf}}$ is the relative humidity at the surface, assumed to be 0.8, and $p_{\mathrm{surf}}$ the pressure at the surface. In the stratosphere the water vapor mixing ratio is set constant to the value at the cold trap.

\subsection{Scenarios}
\label{sec:runs}
We conducted three sets of model scenarios (see Table \ref{tab:runs}). In the nominal scenarios, we vary the relative humidity, surface albedo, and stellar insolation of Earth-like planets around M,K,G, and F-type stars to evaluate the influence of relative humidity and surface albedo upon surface temperatures and to compare 1D modeling results with the growing set of 3D model results in the literature. We chose two additional sets of scenarios for a more detailed comparison with the 3D model results by \cite{Kunze2014} and \cite{Godolt2015} to estimate whether and under which assumptions the global mean surface temperatures calculated by the 3D models may be reproduced by the 1D model.

All of the model calculations have been carried out for Earth-like planets around M, K, G, and F-type main-sequence stars. The planet is assumed to have the same mass and radius as the Earth. As the model calculates diurnally global mean temperature and water vapor profiles, it assumes that the incident stellar radiation is evenly distributed. The composition of the planetary atmosphere is assumed to be Earth-like, i.e.,~a 1bar, N$_2$-dominated atmosphere with 21\% O$_2$, 355ppm CO$_2$, a present Earth O$_3$ profile, 1.64ppm CH$_4$, and  1\% argon. The amount of water vapor is calculated from the atmospheric temperature profile assuming a RH profile as described above. 

For the nominal scenarios, planets with different surface albedos ($A_{\mathrm{surf}}$) are assumed to account for the wide range of possible surfaces of a planet with a water covered surface, from 0.07, a completely ice-free, Earth-like ocean, and 0.8, a completely frozen ocean with snow cover. A planet completely covered with ice cannot be considered habitable in the sense of liquid surface water. However, we chose to include this extreme case to show the possible climate states, hence, also uninhabitable scenarios. Furthermore, one may argue that by increasing the surface albedo in a cloud-free model, one could (partly) account for a net scattering effect by clouds. In between we choose albedos of 0.22, 0.4, and 0.6.

 A surface albedo of $A_{\mathrm{surf}}$=0.22 is the surface albedo needed to obtain the mean surface temperature of Earth for an Earth-like planet around the Sun with the 1D model and an RH profile by \cite{MW1967}. This value is higher than the measured albedo of Earth's surface, which is about 0.13. This is caused by the fact that the 1D model is cloud free, hence the net climatic impact of water clouds is mimicked by an elevated surface albedo. A surface albedo of A$_{\mathrm{surf}}$=0.4 is close to the surface albedo found for \cite{Kunze2014} for their coldest habitable planetary scenario (Ax3.5CO2). A surface albedo of $A_{\mathrm{surf}}$=0.6 is the global mean surface albedo obtained by \cite{Godolt2015} for their uninhabitable planetary scenario (F3D glaciated).\\ 

We varied the orbital distances of the planet, and hence the top of the atmosphere stellar insolation (S) of the M, K, G, and F-type stars,  to obtain  surface temperatures between about \unit[200]{K} and \unit[340]{K}. We selected these temperatures to evaluate the possible climate states resulting from the assumption of different relative humidities and surface albedos. We chose \unit[340]{K} as an upper temperature limit as Earth-like planets may undergo water loss for higher temperatures leading to uninhabitable surface conditions \citep[see, e.g.,][]{Wolf2015,Selsis2007}.\\

The incident stellar spectra are composite spectra of stellar model spectra and measurements for ADLeo (M-Star), $\epsilon$ Eri (K-Star), the Sun (G-Star), and $\sigma$ Boo (F-Star), as described in \cite{Kitzmann2010}. The effective temperatures of these stars are \unit[3400]{K} (M-dwarf), \unit[5072]{K} (K-dwarf), \unit[5777]{K} (Sun), and \unit[6722]{K} (F-dwarf).  \\

For a more detailed comparison with sample 3D model calculations by \cite{Kunze2014} and \cite{Godolt2015}, which are discussed in Sec.~\ref{sec:3d_comp}, the stellar insolation (S) was set to that assumed in these studies. In addition to the above-mentioned values, the surface albedo was set to the global mean surface albedos of the 3D model calculations (A3D). 

For the comparison with the early Earth scenarios with anoxic atmospheres by \cite{Kunze2014},  simple N$_2$, CO$_2$, H$_2$O atmospheres are assumed (see sec.~\ref{sec:3d_comp}) applying the radiative transfer described by \cite{vonParis2010}.

For the scenarios by \cite{Godolt2015}, the influence of the relative humidity profiles by \cite{Cess1976} and \cite{Kasting1986} and a constant relative humidity of 80\%  (RH 80) was investigated.
Furthermore, the relative humidity profiles resulting from the 3D model calculations were parametrized and included in the 1D model. See Sec.~\ref{sec:3d_comp} for details.

\begin{table*}[ht!]
\begin{center}
\begin{tabular}{l|l|l|l|l|l}
Scenario                                        & RH                                    & $A_{\mathrm{surf}}$      & Stars & Atmosphere & S (S$_{\mathrm{Sun}}$) \\ \hline\hline
RH 100 A 0.07                   & RH 100                                &  0.07                                   & \multirow{7}{*}{M, K, G, F-dwarf} & \multirow{7}{*}{Earth-like} & \multirow{7}{*}{yielding 200K$\leq$T$_{surf}\leq$ 340K} \\
RH 100 A 0.22                   &RH 100                                         & 0.22                                    & &&\\
RH MW A 0.07                    & RH MW                         & 0.07                                  &  &&\\
RH MW A 0.22                    &RH MW                          & 0.22                                  &&&\\
RH MW A 0.4                             &RH MW                          & 0.4                                     & &&\\
RH MW A 0.6                             & RH MW                         & 0.6                                     & &&\\
RH MW A 0.8                             & RH MW                         & 0.8                                     &&&\\
\multicolumn{6}{c}{For comparison with \cite{Kunze2014}}\\
\multirow{2}{*}{RH 100 A 0.07}& \multirow{2}{*}{RH 100} &  \multirow{2}{*}{0.07} & \multirow{12}{*}{G-dwarf} & N$_2$, H$_2$O, CO$_2$ (367ppm) & 1, 0.82\\
                                        &                        &                        &                          & N$_2$, H$_2$O, CO$_2$ (3670ppm)& 0.82, 0.77\\
\multirow{2}{*}{RH MW A 0.22} & \multirow{2}{*}{RH MW} & \multirow{2}{*}{0.22}  &  & N$_2$, H$_2$O, CO$_2$ (367ppm) & 1, 0.82\\
                                         &                        &                        &                          & N$_2$, H$_2$O, CO$_2$ (3670ppm)& 0.82, 0.77\\
\multirow{2}{*}{RH MW A 0.4} & \multirow{2}{*}{RH MW} & \multirow{2}{*}{0.4}          &  & N$_2$, H$_2$O, CO$_2$ (367ppm)& 1, 0.82\\
                                        &                        &                        &                          & N$_2$, H$_2$O, CO$_2$ (3670ppm)& 0.82, 0.77\\
\multirow{2}{*}{RH MW A 0.8} & \multirow{2}{*}{RH MW} & \multirow{2}{*}{0.8}          &  & N$_2$, H$_2$O, CO$_2$ (367ppm) & 1, 0.82\\
                                        &                        &                        &                          & N$_2$, H$_2$O, CO$_2$ (3670ppm)& 0.82, 0.77\\
\multirow{2}{*}{RH 100 A3D}& \multirow{2}{*}{RH 100}  & 0.12, 0.75               & & N$_2$, H$_2$O, CO$_2$ (367ppm)& 1, 0.82\\
                                                & &0.29, 0.34           &  & N$_2$, H$_2$O, CO$_2$ (3670ppm) & 0.82, 0.77 \\
\multirow{2}{*}{RH MW A3D} & \multirow{2}{*}{RH MW} & 0.12, 0.75        & & N$_2$, H$_2$O, CO$_2$ (367ppm)& 1, 0.82 \\
                                                & &0.29, 0.34            & & N$_2$, H$_2$O, CO$_2$ (3670ppm) & 0.82, 0.77 \\
\multicolumn{6}{c}{For comparison with \cite{Godolt2015}}\\
RH MW A3D                               & RH MW                 & 0.1, 0.15, 0.21, 0.6 & K, G, F-dwarf & Earth-like & 1\\
RH 100 A3D                              & RH 100                                & 0.1, 0.15, 0.21, 0.6 & K, G, F-dwarf & Earth-like & 1\\
RH KA A 0.22                            & RH KA                         & 0.22                    & K, G, F-dwarf & Earth-like & 1\\
RH Cess A 0.22                  & RH Cess                       & 0.22                  & K, G, F-dwarf & Earth-like & 1\\
RH 3D1 A 0.22                   & RH 3D1                        & 0.22                  & K, G, F-dwarf & Earth-like & 1\\
RH 3D1 A3D                              & RH 3D1                        & 0.1, 0.15, 0.21, 0.6 & K, G, F-dwarf & Earth-like & 1\\
RH 3D2 A 0.22                   & RH 3D2                        & 0.22                  & K, G, F-dwarf & Earth-like & 1\\
RH 3D2 A3D                              & RH 3D2                        & 0.1, 0.15, 0.21, 0.6 & K, G, F-dwarf & Earth-like & 1\\

\end{tabular}
\end{center}
\caption{Scenarios assumed for the 1D model calculations.}
\label{tab:runs}
\end{table*}

\section{Results}

\begin{figure}
\begin{center}
\includegraphics[width=0.49\textwidth]{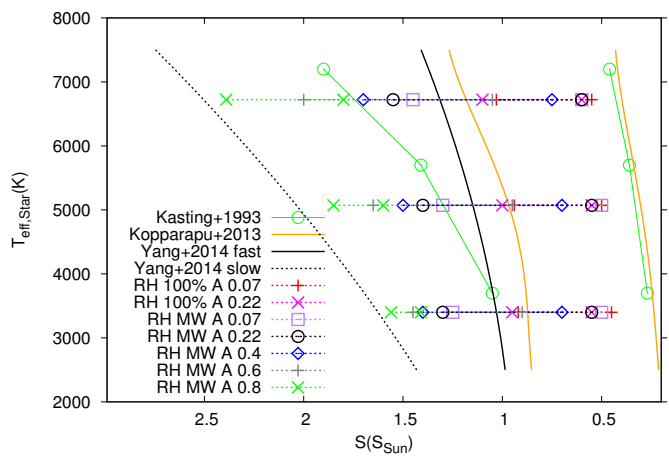}
 \caption{Range of stellar insolations (S) for which 1D model scenarios result in habitable surface temperatures between about \unit[235]{} and \unit[340]{K} compared to HZ boundaries by \cite{kasting1993hz}, \cite{Kopparapu2013}, and \cite{Yang2014} for fast and slow planetary rotation.}
 \label{fig:1DHZ}
\end{center}
\end{figure}

Figure \ref{fig:1DHZ} shows where the 1D model scenarios leading to habitable surface conditions are located in comparison to the habitable zone boundaries as calculated by \cite{kasting1993hz}, \cite{Kopparapu2013}, and \cite{Yang2014} in terms of stellar insolations S. 
Here we consider scenarios as habitable if the surface temperatures calculated by the 1D model fall between \unit[235]{K} and \unit[340]{K}. We chose \unit[235]{K} as the lower temperature boundary, as 3D model calculations have shown that habitable regions with liquid surface water may exist on planets that have global mean surface temperatures below \unit[273.15]{K}, see, for example, \cite{Charnay2013}, \cite{Wolf2013}, \cite{Kunze2014}, and \cite{Shields2014}. The lowest global mean surface temperature for a planet that still shows liquid surface water obtained by these 3D model calculations is about \unit[235]{K} for an Earth-like planet around an M-dwarf star by \cite{Shields2014}. Below this temperature we assume that the planet is uninhabitable because of global glaciation. The upper temperature limit has been chosen according to the results by, for example,~\cite{Wolf2015} and \cite{Selsis2007}, which show that at temperatures higher than about \unit[340]{K} water loss to space becomes 
important and may render the planet uninhabitable. The 
HZ boundaries of the studies by \cite{kasting1993hz} and \cite{Kopparapu2013} are the so-called runaway greenhouse limit for the inner HZ and the so-called maximum greenhouse limit for the outer HZ. The inner HZ boundaries by \cite{Yang2014} are those determined for slowly and rapidly rotating Earth-like planets.

The habitable scenarios extend over a wide range of orbits within the HZ and even beyond, depending on which HZ boundary study is referred to. The only inner HZ boundary that is not reached by the 1D model calculations is the boundary computed by \cite{Yang2014}, for slowly rotating Earth-like planets, using a 3D climate model. They found that slowly rotating planets with Earth-like atmospheres may build up a cloud layer on the star-lit side, which strongly increases the planetary albedo and thereby allows for smaller orbital distances. This negative cloud feedback was recently confirmed by \cite{Kopparapu2016}. The 1D modeling scenarios leading to the smallest possible orbital distances are those with an Earth-like relative humidity profile and the high surface albedo of 0.8, as expected. Even for this combination of relative humidity and surface albedo, the very high planetary albedo obtained by \cite{Yang2014} for the slowly rotating planets cannot be reproduced as the water vapor in the atmosphere partly 
masks the surface albedo; this leads to planetary albedos that are smaller than those obtained by the 3D model study at these relatively high temperatures above \unit[300]{K}.

None of the 1D model scenarios lies outside the outer edge of the HZ since obtaining habitable conditions at such low stellar insolations requires a higher amount of greenhouse gases such as CO$_2$, which we did not increase in our model calculations presented in Fig.~\ref{fig:1DHZ}. We do not expect that the assumption of different surface albedos and relative humidities would have a big impact on the outer edge of the habitable zone since water vapor concentrations would be very low owing to low surface temperatures. This would lead to low water vapor saturation pressures, regardless of which relative humidity profile is assumed (see below). Furthermore, for the variation in surface albedo we expect a smaller impact as for the scenarios discussed here, since, to reach habitable surface temperatures at the outer edge of the habitable zone, high amounts of CO$_2$ are required that mask the surface albedo, as shown by \cite{Shields2013} and \cite{vonParis2013ice}. As expected, the largest orbital 
distances for habitable surface conditions are found for the Earth-like planet when assuming the lowest surface albedo of 0.07 and a high relative humidity (fully saturated atmosphere, corresponding to the highest amount of greenhouse gases in the considered scenarios).

\subsection{Influence of relative humidity and surface albedo on 1D model results}
\label{sec:RH_alb}

\begin{figure*}
\begin{center}
\includegraphics[width=0.49\textwidth]{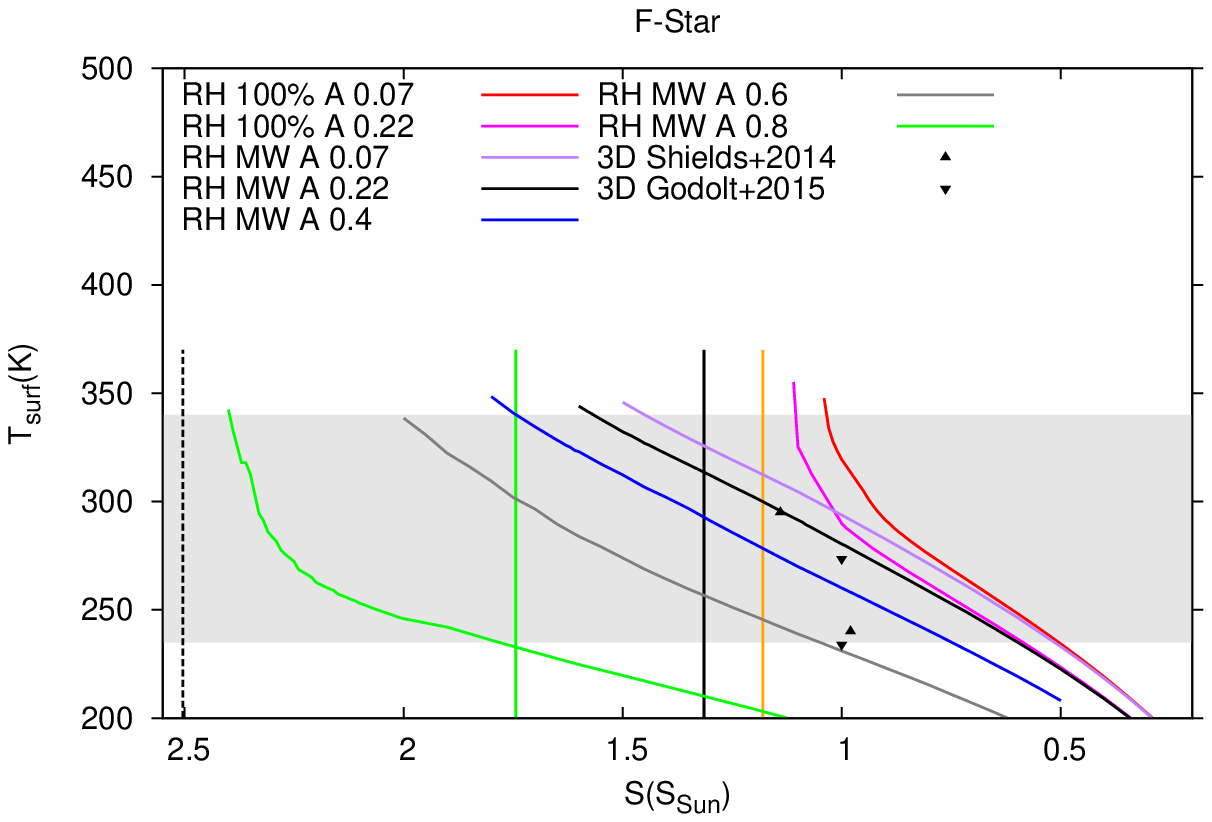}
\includegraphics[width=0.49\textwidth]{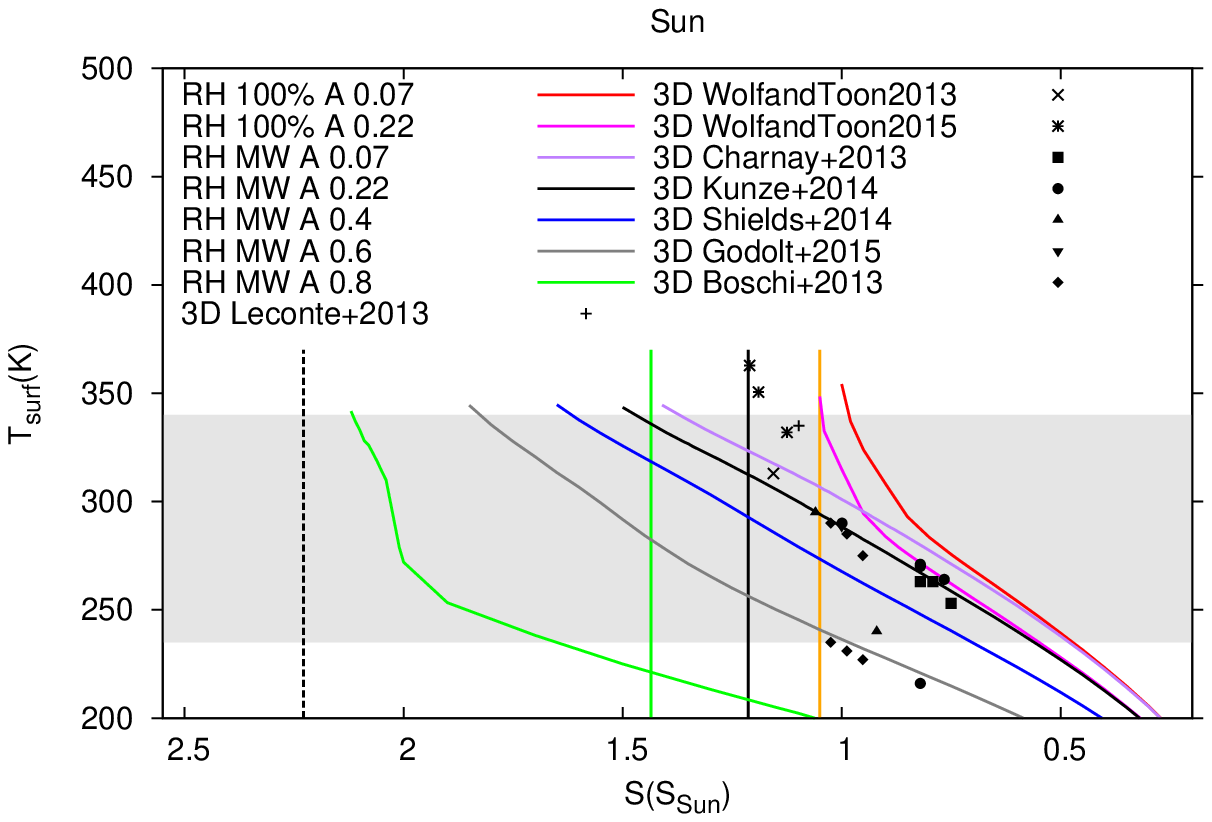}
\includegraphics[width=0.49\textwidth]{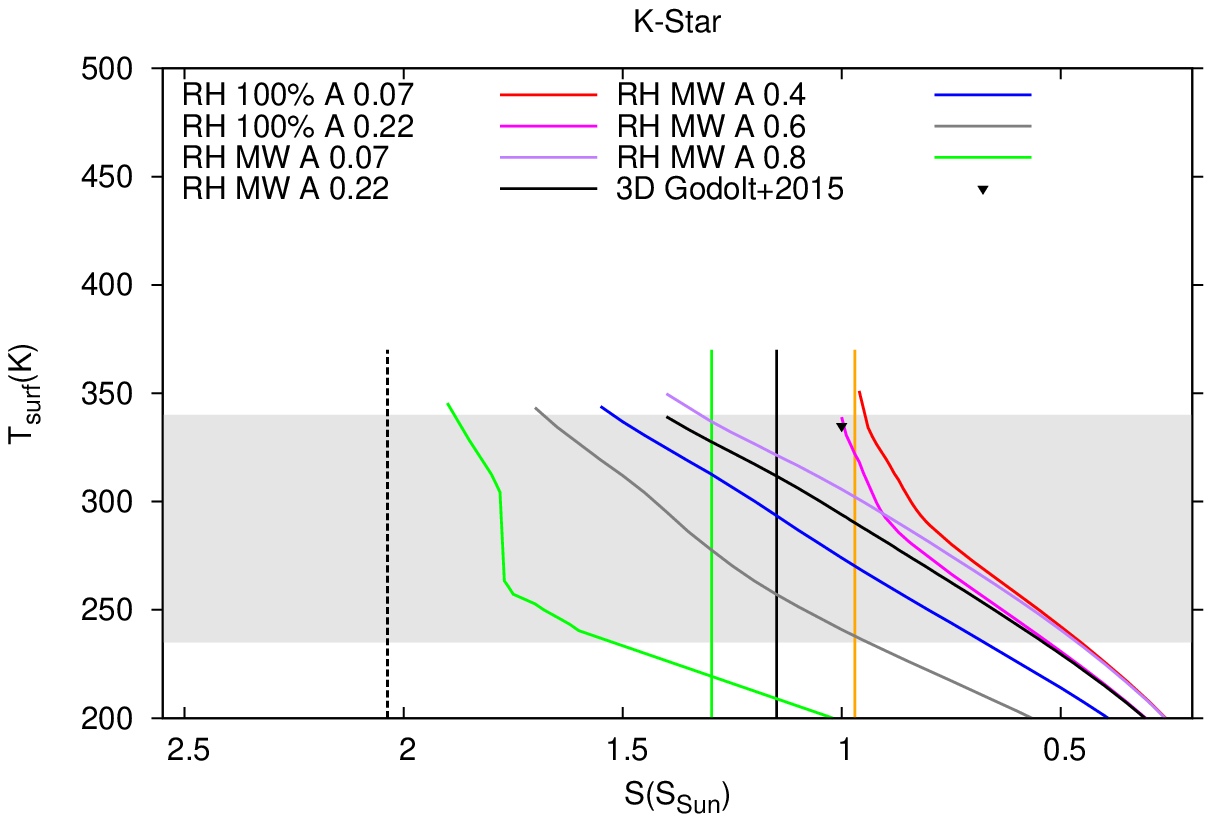}
\includegraphics[width=0.49\textwidth]{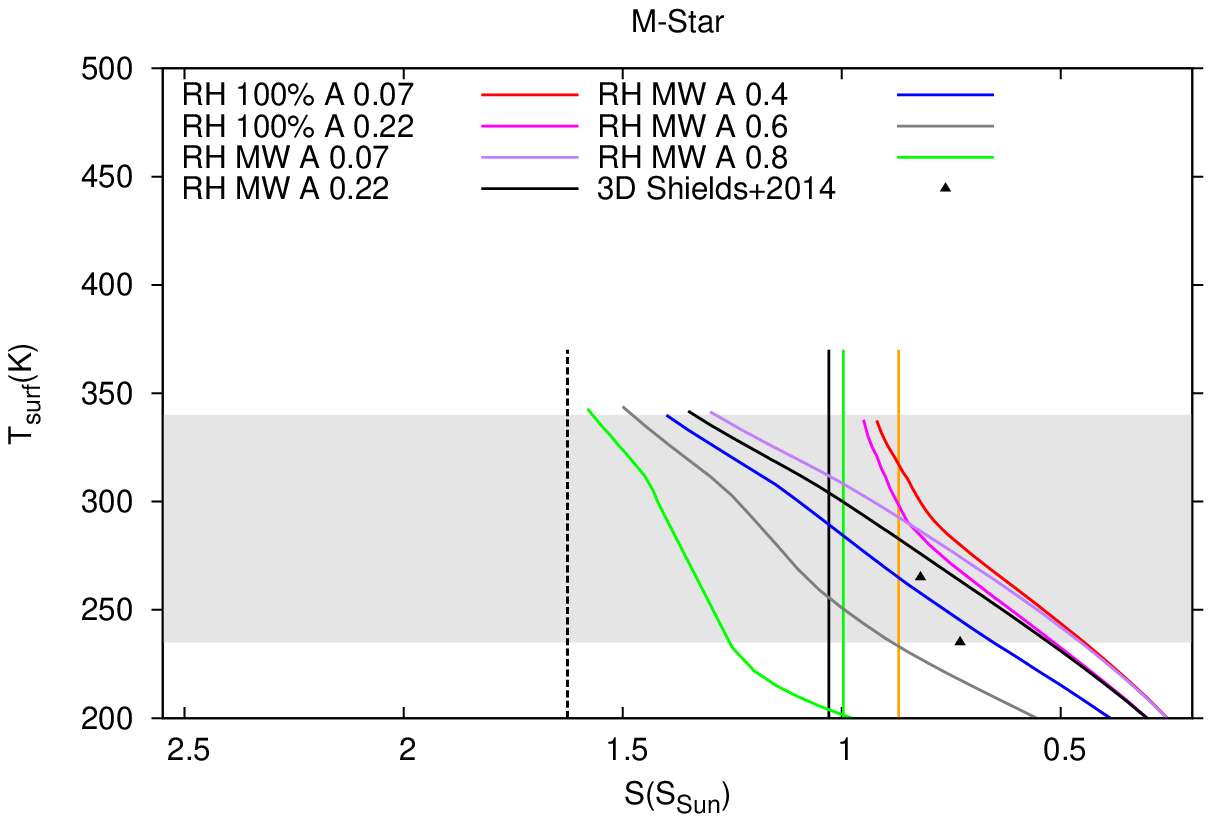}
\end{center}
\caption{Surface temperatures of Earth-like planets around different stars at various stellar insolations \textcolor{black}{(S)} as computed with the 1D model for different assumptions of the relative humidity and surface albedo. Upper left panel: Earth-like planets around an F-type star; upper right panel: Earth-like planets around the Sun; lower left panel: Earth-like planets around a K-type star; and lower right panel: Earth-like planets around a M-type star. The vertical lines indicate the inner habitable zone boundaries as calculated by \cite{Kopparapu2013} (orange), \cite{kasting1993hz} (green), and \cite{Yang2014} (solid black line for rapidly rotating planets and dashed black line for slowly rotating planets). Symbols indicate the results of 3D modeling studies by \cite{Boschi2013}, \cite{Yang2013}, \cite{Charnay2013}, \cite{Godolt2015}, \cite{Kunze2014}, \cite{Leconte2013}, \cite{Shields2014}, \cite{Wolf2015}, and \cite{Wolf2013}. Some of the 3D modeling scenarios of the early Earth \citep{
Charnay2013,Kunze2014} have larger CO$_2$ volume mixing ratios than assumed for Earth-like planets. }
\label{fig:ToverS}
\end{figure*}

The assumptions of different relative humidities and surface albedos in the 1D model calculations lead to a wide range of surface temperatures for one specific stellar insolation. For example, for an Earth-like planet located at \unit[1]{au} around the Sun (with a stellar insolation of 1 solar constant (S$_{\mathrm{Sun}}$)), the surface temperatures (T$_{surf}$) range from \unit[196.4]{K} to \unit[354.2]{K}; see Table \ref{tab:surfT_GS1}. Hence, according to our 1D model calculations, at \unit[1]{au} an Earth-like planet could show habitable and uninhabitable (too cold, and too warm) climate states depending on the surface properties, the behavior of the water vapor feedback, and water loss to space.

\begin{table}
\begin{center}
\begin{tabular}{l|l|c}
Scenario 
& T$_{\mathrm{surf}}$ (K) & Climate state\\ \hline
RH 100 A 0.07&  354.2 & water loss ($T_{\mathrm{surf}}>$\unit[340]{K})\\
RH 100 A0.22 &  314.9 &  habitable\\
RH MW A0.07 &  301.0 & habitable\\
RH MW A0.22 & 288.3 & habitable\\
RH MW A0.4 &  267.7 & habitable\\
RH MW A0.6 &  236.4 & habitable\\
RH MW A0.8 &  196.4 & glaciation ($T_{\mathrm{surf}}<$\unit[235]{K})\\
\end{tabular}
\end{center}
\caption{Resulting surface temperatures (T$_{\mathrm{surf}}$)for an Earth-like planet at \unit[1]{au} around the Sun for different assumptions of relative humidity and surface albedo in the 1D model calculations. The possible climate states (uninhabitable because of water loss, habitable, or uninhabitable because of glaciation) have been determined from the resulting surface temperatures.}
\label{tab:surfT_GS1}
\end{table}

Figure \ref{fig:ToverS} shows the surface temperatures of Earth-like planets at various stellar insolations S around different host stars assuming different relative humidity profiles and surface albedos. An increase in stellar insolation corresponds to a decrease in orbital distance and leads to an increase of the planetary surface temperature for all scenarios, as expected.

Upon increasing the surface albedo, the stellar incident radiation needed to obtain the same surface temperature also increases. Planets with a higher surface albedo can be closer to the star and still remain habitable.
The difference in surface temperature as a result of different surface albedos slightly decreases with increasing surface temperature. This is a result of the higher optical depth of the atmosphere due to increased water vapor for higher temperatures, which partly masks the surface albedo. This effect is most prominent for the planetary scenarios around an M-dwarf star (see lower right panel of Fig.~\ref{fig:ToverS}), since for these scenarios the absorption of stellar light by water vapor is enhanced because water vapor mainly absorbs the near-infrared part of the stellar flux, which is larger for M-dwarf stars than, for example, for~G-type stars.\\

The different relative humidity parameterizations show a stronger impact on the surface temperatures when surface temperatures are high and the atmospheres can contain more water vapor. For planets with surface temperatures above about \unit[300]{K}, temperatures increase nearly linearly with stellar incident radiation for the Earth-like relative humidity profile (indicated by RH MW), while for the assumption of a saturated atmosphere (indicated by RH 100) the temperature increase is steeper from the strong increase in water vapor following the increase in water saturation vapor pressure with temperature.\\

For a surface albedo of 0.8 the surface temperature increase with stellar incident radiation shows different behaviors for low (below about 250K) and high temperatures, which are both different from the increase in temperature for the other surface albedos. For low temperatures, there is only little water vapor in the atmosphere and only a little stellar radiation is absorbed in the atmospheres. For these low temperatures, the surface temperature increase with stellar insolation is less that what is observed for the lower surface albedos since for the high surface albedo only a very small part of the stellar radiation reaching the surface can be used to heat the surface because most of it is reflected. At higher surface temperatures, however, more water vapor is present in the atmospheres, which partly masks the high surface albedo. This leads to an increase in the lower atmospheric temperatures and, thereby, also in the greenhouse effect. This effect would actually be weaker for planets around cooler stars, 
as calculated here, if the high albedo was caused by snow and ice because the surface albedo of ice and snow is wavelength dependent and lower at longer wavelengths, where cooler stars have their radiation 
maximum \citep[see, e.g.,][]{Shields2013,Joshi2012}. \\

It can be inferred from our 1D modeling results that depending on surface albedo and relative humidity very different surface temperatures could result for the same stellar insolation.  For example, for a planet around a G-type star, with a solar insolation (S) of about 1.2, the planet could be completely frozen with temperatures around \unit[200]{K} (as obtained for a surface albedo of 0.8 and an Earth-like relative humidity), habitable with surface temperatures between 290 and about \unit[320]{K}, or uninhabitable because of very high surface temperatures (as would be obtained for a relative humidity of 100\%, similar to the estimates by \cite{Kopparapu2013}). See also Table \ref{tab:surfT_GS1}. 

This degeneracy was also discussed in the context of 3D modeling studies \citep[see, e.g.,][]{Boschi2013,Shields2014}, where different climate states and surface temperatures were obtained despite the interactive calculation of the sea-ice evolution and water vapor feedback. The climate state obtained in their studies primarily depends on the initial conditions of the model, in particular, the climate state via the initial water vapor and surface albedo. A degeneracy was also found in the 3D modeling study of \cite{Godolt2015}. In their study, different assumptions for the oceanic heat flux have led to different climate states (habitable or glaciated). Also the 3D modeling study of \cite{Yang2013} finds different surface temperatures at a constant solar insolation with temperature differences as large as \unit[56]{K} depending on which oceanic heat transport was assumed. This demonstrates that, even though 3D models are able to account for the water vapor and ice albedo feedback when calculating the 
hydrological cycle and build-up of snow and sea ice, it is still challenging to judge which surface temperatures would be realized; this is because how the hydrological cycle and the ice albedo respond depends on various factors. Exploring the full parameter range of possible solutions for a given planet with different atmospheric mass and composition by utilizing a 1D model thereby gives a first good estimate of which surface temperatures could be realized on such a planet when including the range of possible surface albedos and relative humidities. Which surface temperatures would be realistic under given assumptions and how climate feedbacks, such as the water vapor, ice albedo and cloud feedback, operate and interact, can however only be assessed by detailed 3D climate modeling, which treat these processes self-consistently.

While the degeneracy from the initial state or different oceanic heat transport found in 3D models can possibly be captured by parameter studies of relative humidity and surface albedo such as this study, the degeneracy found by \cite{Yang2014} for planets with a slow and a fast rotation rate may be hard to overcome. These authors found that clouds may build up on the day side of a slowly rotating planet, reflecting a large part of the stellar incident radiation so that such planets can have habitable surface temperatures closer to the star than predicted for relatively rapid rotating planets (as the Earth). Since this effect is caused by clouds, it cannot be captured in our 1D cloud-free model. The high cloud reflectivity may only be partly approximated by increasing the surface albedo in our 1D model. With the assumption of a high surface albedo of 0.8 and an Earth-like relative humidity, we find a planetary albedo of about 0.41 for the Earth-like planet around the G-type star at \unit[340]{K} surface 
temperature. This is low compared to the result by \cite{Yang2014} who reported a value of about 0.64 for the slowly rotating planet around a G-type star. While \cite{Yang2014} obtain an increasing planetary albedo with increasing stellar insolation due to the buildup of clouds, the planetary albedo of our 1D model calculations decreases with increasing the stellar insolation since for higher surface temperatures, water vapor in the atmosphere increases and the surface albedo is masked more efficiently. The tendency of thick atmospheres to mask the surface albedo has been discussed for CO$_2$-dominated atmospheres by, for example, \cite{Shields2013} and \cite{vonParis2013ice}. Hence, to estimate the inner boundary of the habitable zone, where temperatures are higher and, hence, atmospheric absorption from water vapor is also higher, a cloud-free model as used here cannot approximate such a negative cloud-feedback on surface temperatures by increasing the surface albedo. It should be kept in mind 
that the strengths of the cloud albedo effect found by \cite{Yang2014} decreases for increasing efficiency of the oceanic heat transport, as shown in \cite{Yang2013}. Furthermore, quantifying the cloud feedback is a great challenge even in Earth climate research  \citep[see, e.g.,][]{flato2013}, so more work is required to estimate the range of possible cloud feedbacks.\\
 
Figure \ref{fig:ToverS} shows the 1D model calculations conducted here and the results of 3D model calculations from the literature. For all four stellar types, most of the planetary surface temperatures calculated with the 3D models lie within the range of the 1D model results. Hence,  the 1D model can approximate the resulting surface temperature with a certain set of relative humidity and surface albedo for these 3D
model calculations. 
The only 3D model result for which the 1D model has difficulty approximating the results is that by \cite{Yang2014} for slowly rotating planets, as mentioned above. The results by \cite{Yang2014} for rapidly rotating planets can however be approximated by the 1D model. Most of the habitable 3D model scenarios lie in between the 1D model calculations, with a surface albedo of 0.4 and a relative humidity profile of \cite{MW1967}, and calculations with a fully saturated atmosphere and a surface albedo of 0.22. Exceptions to this, apart from the \cite{Yang2014} slowly rotating planets, are the 3D model results by \cite {Shields2014} with warm initial climate conditions. For \cite{Shields2014}, we show the first and last habitable scenario, after and before global glaciation, depending on whether the model was started from a cold or a warm climate state, respectively. 
Their warm start scenarios have very low surface temperatures, as low as \unit[235]{K}, which would usually not be considered to be habitable when utilizing a 1D model to estimate the habitability of a planet via its global mean surface temperature.\\

However, the resulting surface temperatures by \cite{Shields2014}, as well as those of the  non-habitable glaciated 3D model scenarios by \cite{Boschi2013}, \cite{Kunze2014}, and \cite{Godolt2015} can be approximated in 1D by assuming a relative humidity of \cite{MW1967} and a surface albedo of around 0.6.
\\

Hence, when treating the surface albedo and relative humidity profile as parameters in 1D model studies and using the habitability constraints as found by recent 3D modeling studies, the same conclusions about the potential habitability of a planet can be drawn as from 3D model calculations.

In the following section we give a more detailed comparison of the 3D model results by \cite{Kunze2014} and \cite{Godolt2015}.

\subsection{Comparison of 1D model calculations with 3D model results}
\label{sec:3d_comp}

The water vapor and ice albedo feedback enhance the climatic response to changes, for example, in stellar insolation. These processes are usually not captured by 1D models but are part of most state-of-the-art 3D climate models. We investigate here which assumptions of the parameters representing different realizations of water vapor distribution, hence relative humidity profiles and surface albedos, yield similar results as the 3D model calculations performed by \cite{Kunze2014} and \cite{Godolt2015}.
These 3D model scenarios represent a wide range of possible planetary climates. While \cite{Kunze2014} found habitable scenarios at global mean temperatures, which would usually not be considered habitable as they are below \unit[273]{K}, \cite{Godolt2015} found a wide range of surface temperatures of habitable scenarios along with an uninhabitable, glaciated scenario.\\

\begin{table}
\begin{center}
\begin{tabular}{l|p{1.5cm}|p{1.5cm}|l|l}
Scenario & Stellar Insolation at TOA (W/m$^2$)& vmr CO$_2$ (ppm) & T$_{\mathrm{surf}}$ (K) &A$_{\mathrm{surf}}$\\ \hline
Ax&           1365  &  367  &      290.35  &  0.12      \\ 
Ax2.5 &      1121  & 367    &     216.05  &  0.75       \\   
Ax2.5CO2 & 1121  & 3670   &    270.45  &  0.29       \\   
Ax3.5CO2 & 1046   & 3670    &   264.45  &  0.34      \\      
\end{tabular}
\end{center}
\caption{Summary of model scenarios by \cite{Kunze2014} of anoxic (Ax) atmospheres at different ages of the Earth (present, 2.5 and \unit[3.5]{Ga} ago) used for the comparison with 1D model results.}
\label{tab:Kunze}
\end{table}

\begin{figure}
\begin{center}
\includegraphics[width=0.5\textwidth]{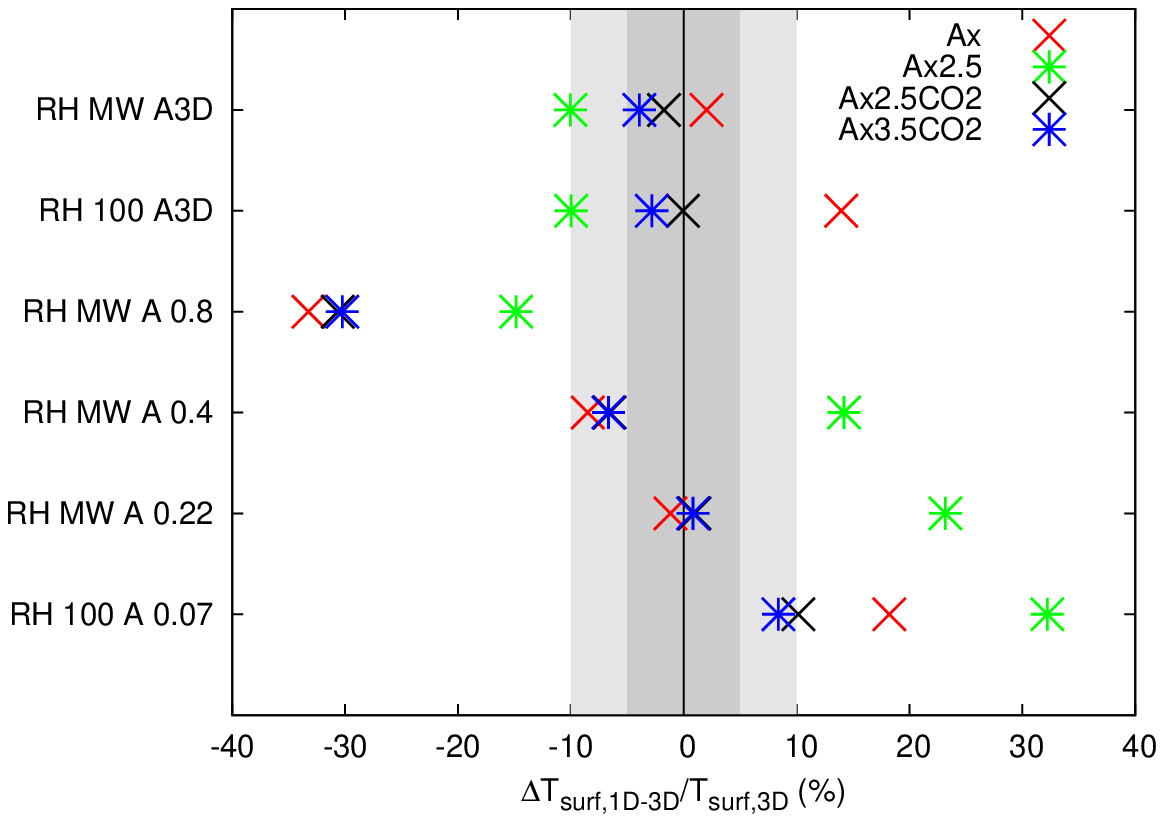}
 \caption{Deviation of surface temperatures calculated by the 1D model from the surface temperatures calculated by a 3D model $\left(\frac{T_{surf,1D}-T_{surf,3D}}{T_{surf,1D}}\right)$ by \cite{Kunze2014} in percent.  The gray shaded areas indicate differences of  5 and \unit[10]{\%}. }
 \label{fig:Kunze}
\end{center}
\end{figure}

\cite{Kunze2014} investigated the climate of the early Earth during the Archean eon with a 3D climate model \citep[EMAC - ECHAM/MESSy Atmosphere Chemistry Model; ][]{Joeckel2006} by assuming aqua planets with anoxic atmospheres. The total stellar insolation for the early Earth at 2.5 and 3.5Ga before present was assumed to have a total energy input at the top of the atmospheres of 82\% and 77\% of the present Sun following \cite{Gough1981}. The stellar spectral distribution was taken from a solar analog star, $\beta$ Com. The CO$_2$ level was increased by a factor of 10 relative to modern Earth's value for
the early Earth scenarios, i.e.,~to \unit[3670]{ppm}, which lead to habitable surface conditions in some locations despite mean surface temperatures below; see Table \ref{tab:Kunze}.\\
We compare these 3D model results with 1D model calculations for similar scenarios. Since \cite{Kunze2014} found that the spectral stellar flux distribution did not strongly influence the surface temperatures for these anoxic atmospheres, we assume the stellar spectrum of the Sun instead of $\beta$ Com. We furthermore assume that the atmospheres only consist of N$_2$, CO$_2$, and H$_2$O, and thereby neglect the greenhouse effect of methane, which was included in the 3D model calculations at a concentration of \unit[1.7]{ppm}. Table \ref{tab:Kunze} gives details of the 3D model scenarios and resulting surface temperatures and surface albedos.

In Fig.~\ref{fig:Kunze} the 3D model results are compared to those of the 1D model. We assumed different parameters sets of relative humidity and surface albedo for the 1D model calculations that are similar to the results presented in Sec.~\ref{sec:RH_alb}. The comparison shows that for the 3D model scenarios with habitable surface conditions (Ax, Ax2.5CO2, Ax3.5CO2), the surface temperature is  approximated well by assuming the relative humidity profile by \cite{MW1967} and a surface albedo of 0.22. For the coldest scenario, Ax2.5, which is not habitable as the entire surface is covered with sea ice, a surface albedo of 0.22 leads to higher surface temperatures than those found by the 3D model calculations. This is also the case for a surface albedo of 0.4. For a surface albedo of 0.8, however the temperatures are too low, hence an albedo between 0.4 and 0.8 would lead to a good approximation.

When assuming the surface albedos resulting from the 3D model calculations as an input to the 1D model calculations, we find an overall acceptable approximation of the surface temperatures for the habitable scenarios when assuming a relative humidity of the present Earth. This leads to a slight underestimation of the 3D global mean surface temperature by the 1D model calculations for the Ax2.5CO2 and the Ax3.5CO2 scenarios. An overestimation of the 3D model result by the 1D model is found for the scenario with present Earth insolation (Ax).  The surface temperature is underestimated by about \unit[10]{\% for the Ax2.5 scenario}. An explanation for this is that in the 3D model clouds increase the greenhouse effect of the atmospheres for the cold scenarios (Ax2.5, Ax2.5CO2, Ax3.5CO2), while they increase the scattering effect for the warmer scenario (Ax). \cite{Kunze2014} found a more negative cloud radiative forcing for the 
Ax scenario than for present Earth, while they found a less negative cloud radiative forcing for colder scenarios, which is 
nearly neutral for Ax3.5CO2 and slightly positive for Ax2.5 (see their Table 4). \\

While for the colder scenarios (Ax2.5, Ax2.5CO2, Ax3.5CO2) a similarly good approximation of the global mean climate of the 3D model calculations by the 1D model can be found when using the 3D model surface albedo and the assumption of a fully saturated atmosphere, the discrepancy is larger for the warmer scenario with present solar irradiation (Ax). For this scenario, assuming a relative humidity of 100\%  overestimates the global mean climate of the planet by about \unit[30]{K} in the 1D model calculations. This is caused by the fact that, for the colder scenarios, the water vapor feedback is weak, as there is only a little water in the atmospheres regardless of the relative humidity profile assumed because the water vapor saturation pressure is low.  For the scenario with present solar irradiation (Ax), however, the greenhouse effect of water vapor and hence the water vapor feedback cycle, is more important because surface temperatures are higher and overestimated by the assumption of a fully saturated 
atmosphere. \\ 

For the habitable scenarios in \cite{Kunze2014}, it can be concluded that the assumption of a surface albedo of 0.22 and the Earth-like relative humidity profile \citep{MW1967} in the 1D model best approximates the global mean surface temperatures of these 3D climate calculations for habitable scenarios. This set of surface albedo and relative humidity is assumed in 1D climate-chemistry calculations of, for example, \cite{Segura2003,Segura2005}, \cite{Rauer2011}, and \cite{Grenfell2014}.\\

\begin{table}
\begin{center}
\begin{tabular}{l|p{1.cm}|l|l|p{1.5cm}}
Scenario &  Stellar Type & T$_{\mathrm{surf}}$ (K) &A$_{\mathrm{surf}}$ & Comment\\ \hline
F3D glaciated &      F-type star  & 234.0    &  0.6  &   no oceanic heat redistribution     \\   
F3D&        F-type star  &  273.6  &      0.21  &        \\ 
G3D&  G-type star   &   288.6   &  0.15  &     \\    
K3D   & K-type star    &   334.9  &  0.1    &  \\     
\end{tabular}
\end{center}
\caption{Summary of model scenarios by \cite{Godolt2015} used for the comparison with 1D model results.}
\label{tab:Godolt}
\end{table}

\begin{figure}
\begin{center}
\includegraphics[width=0.5\textwidth]{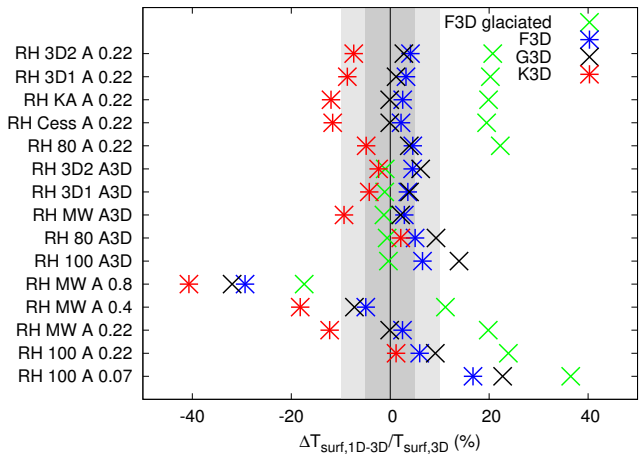}
 \caption{Deviation of surface temperatures calculated by the 1D model from the surface temperatures calculated by a 3D model (as in Fig.~\ref{fig:Kunze}) by \cite{Godolt2015} in percent.  The gray shaded areas indicate differences of  5 and \unit[10]{\%}. }
 \label{fig:Godolt}
\end{center}
\end{figure}

A similar result can be found when comparing the 1D model calculations for Earth-like planets around the different types of central stars presented in Sec.~\ref{sec:RH_alb} to the 3D model results \textcolor{black}{of} \cite{Godolt2015}. In their study they modeled Earth-like planets, with Earth-like mass, radius, continents, obliquity, eccentricity, and rotation period at orbital distances where the total stellar irradiation at the top of the atmosphere is equal to Earth's total solar irradiance (or solar constant, $S_{\mathrm{Sun}}$=\unit[1366]{$\tfrac{\mathrm{W}}{\mathrm{m^2}}$}).  They assumed Earth-like atmospheric compositions, hence N$_2$-O$_2$-dominated atmospheres with trace gas amounts of O$_3$, CO$_2$, CH$_4$, and N$_2$O, while H$_2$O was calculated by the 3D climate model.

The results of \cite{Godolt2015} are summarized in Table \ref{tab:Godolt}. These authors found two climate states for the Earth-like planet around the F-type star. Assuming an Earth-like oceanic heat transport led to cold but habitable surface conditions, while neglecting any oceanic heat transport led to global glaciation. Figure \ref{fig:Godolt} shows the surface temperatures of the 3D model calculation of \cite{Godolt2015} in comparison to the 1D model calculations with different sets of relative humidity and surface albedo.

For habitable climates of the Earth-like planet around the G- and F-type star, the 1D model  gives a good approximation of the 3D model results, when assuming a surface albedo of 0.22 and the relative humidity profile of the Earth \citep{MW1967} as also found also for the comparison with the results by \cite{Kunze2014}. For the planet around the K-type star, however, the result can be approximated when assuming a fully saturated atmosphere and a surface albedo of 0.22. This is a rather peculiar result, as the assumption of a fully saturated atmosphere in the 1D model does not give a good approximation to  other 3D model calculations in the literature; see Fig.~\ref{fig:ToverS}. It has furthermore been discussed that fully saturated atmospheres are unrealistic, as atmospheric dynamics tend to dehydrate the air \citep[see, e.g.,][]{Leconte2013b}. The 1D model does not find habitable solutions for a fully saturated atmosphere and surface albedos of 0.07 for the Earth-like planet around the K-type star. 

For a more detailed comparisons of the 3D model results with 1D model calculations, we also used the relative humidity parameterizations by \cite{Cess1976} and \cite{Kasting1986}, which we refer to hereafter as RH Cess and RH KA. Furthermore, we applied a fixed relative humidity of 80\% (RH 80) instead of 100\%. These parametrizations are written as

\begin{equation}
\begin{split}
RH(p)=
\begin{cases}
RH_{surf}\left(\frac{\frac{p}{p_{\mathrm{surf}}}-0.02}{0.98}\right)^{\Omega} \text{with}\\
  \hspace{0.2cm}\Omega=1-0.03(T_{\mathrm{surf}}[K]-288) \rightarrow \text{RH Cess}\\
  \hspace{0.2cm}\Omega=1-\frac{\frac{p_{\mathrm{sat,H_2O,surf}}}{p_{\mathrm{surf}}}-0.0166}{0.0834} \hspace{0.4cm}\rightarrow \text{RH KA}\\
80\% \hspace{3.3cm}\rightarrow \text{RH 80}\\
\end{cases}
\end{split}
.\end{equation}

In addition, the relative humidity profiles resulting from the 3D model calculations were parametrized and included in the 1D model. Here two different approaches were used (RH 3D1 and RH 3D2). Firstly, we followed similar  approaches of the other relative humidity profile parameterizations by \cite{MW1967}, \cite{Cess1976}, and \cite{Kasting1986} and fit the exponent $\Omega_1$ for the four different 3D model scenarios of \cite{Godolt2015}. We then included these fits (RH 3D1) into the 1D model calculations of the corresponding scenarios. Secondly, we built a relative humidity profile, which has a constant relative humidity ($RH_{\mathrm{mean}}$) throughout most of the troposphere up to a pressure of $p_{\mathrm{trop}}$ and above declines with decreasing pressure (RH 3D2). This form of parametrization was chosen as the relative humidity computed by the 3D model follows such a distribution. The two constructed parametrizations can be written as 

\begin{equation}
\begin{split}
RH(p)=
\begin{cases}
RH_{\mathrm{surf}}\left(\frac{p}{p_{\mathrm{surf}}}\right)^{\Omega_1} \rightarrow \text{RH 3D1}\\
   \begin{cases} 
    RH_{\mathrm{mean}}   \text{  for  } p>p_{\mathrm{trop}}\\
    RH_{\mathrm{mean}}\left(\frac{p}{p_{\mathrm{trop}}}\right)^{\Omega_2}   \text{  for  } p<p_{\mathrm{trop}}\\
    \end{cases}
     \rightarrow \text{RH 3D2}\\
\end{cases}
 \end{split}
\label{eq:3DRHS}
  .\end{equation}

\begin{figure}
\begin{center}
\includegraphics[width=0.5\textwidth]{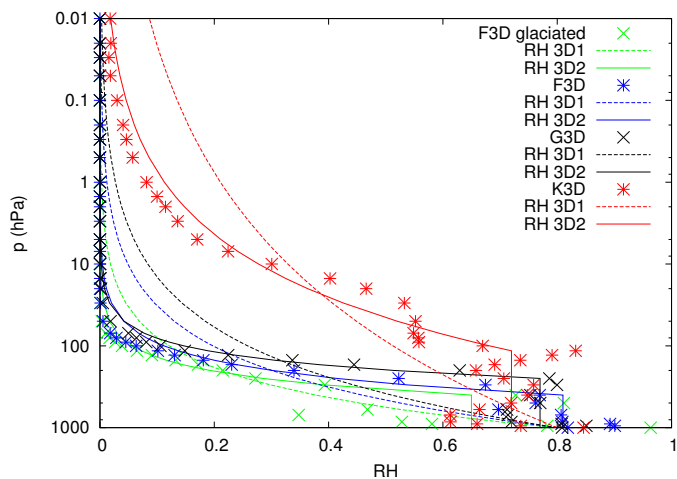}
 \caption{Relative humidities from the 3D model results (symbols) for the four scenarios in \cite{Godolt2015} and their parameterizations RH 3D1 (dashed lines)  and RH 3D2 (solid lines) as given in Eq.~\ref{eq:3DRHS}, which have been included in the 1D model calculations.}
 \label{fig:RHparametrisation}
\end{center}
\end{figure}

Table \ref{tab:RHparameters} gives details about the parameters resulting from the fits of the relative humidity profiles of the 3D model calculations by \cite{Godolt2015} and Fig.~\ref{fig:RHparametrisation} shows the relative humidity parameterizations and the 3D model results.\\

\begin{table}
 \begin{center}
  \begin{tabular}{c|c|c|c|c}
   3D scenario & $\Omega_1$ &  RH$_{\mathrm{mean}}$ & p$_{\mathrm{trop}}$ (bar) & $\Omega_2$\\ \hline
F3D glaciated    &  0.819357  &  0.65       & 0.4              & 1.75615\\   
F3D         &  0.607409  &  0.81        & 0.4              & 1.44578\\
      G3D         &  0.490586  &  0.77       & 0.25             & 1.8076 \\
   K3D         &  0.193146  &  0.72       & 0.115            & 0.390764\\
  \end{tabular}
 \end{center}
 \caption{Parameters derived from the 3D model calculations of \cite{Godolt2015} for the relative humidity parameterizations RH3D1 and RH3D2.}
 \label{tab:RHparameters}
\end{table}

Utilizing the relative humidity parameterizations by \cite{Cess1976} and \cite{Kasting1986} in combination with a surface albedo of 0.22 does not provide any apparent advantage over the relative humidity parametrization by \cite{MW1967} for the cases investigated here.  By assuming a constant relative humidity of 80\% instead of 100\% for the scenario of the
planet around the K-type star, we obtain surface temperatures of \unit[318]{K} and \unit[342]{K} for surface albedos of 0.22 and 0.1 (A3D for this case). Hence, the surface temperature of the Earth-like planet around the K-type star can also be approximated using a relative humidity, which is less than 100\%; this is in better agreement with the modeling study by \cite{Leconte2013} who found that full saturation is prevented by atmospheric dynamics. Furthermore, this agrees better with the results of the 3D model calculations, where the global mean relative humidity in the troposphere is about \unit[72]{\%} for the Earth-like planet around the K-type star.

Using the relative humidity parameterizations constructed from the 3D model results (RH 3D1 and RH 3D2) in the 1D model leads to an overall acceptable agreement of the surface temperatures computed with the 1D and 3D model for the habitable scenarios. The glaciated 3D model scenario of the Earth-like planet around the F-type star can be best approximated by the 1D model when assuming the surface albedo of 0.6 as calculated by the 3D model regardless of the relative humidity assumed. This surface albedo generally leads to a good approximation of the low surface temperatures of glaciated scenarios by the 1D model, as can be seen in Fig.~\ref{fig:ToverS}.

\section{Discussion}
\label{sec:discussion}

\subsection{Surface albedo}
We modeled climates of Earth-like planets for surface albedos between 0.07 and 0.8. This range has been chosen according to the minimum and maximum surface albedo for a planet covered with water, with 0.07 representing ocean water and 0.8 sea ice with snow cover. As the 1D model used here calculated global mean surface temperatures, a global mean surface albedo has to be assumed. When focusing only on habitable scenarios the range of surface albedos is certainly smaller, since a planet with a surface covered by sea ice is not habitable in the sense of availability of liquid water on the planetary surface. Hence, the highest mean surface albedo for habitable conditions would be smaller than 0.8. In the 3D model study of \cite{Kunze2014}, the highest surface albedo obtained for a habitable surface scenario is 0.34 (see Table \ref{tab:Kunze}). However, introducing this surface albedo to the 1D model calculation gives an underestimation of the surface temperature compared with the 3D model result because for 
this scenario the 3D model shows a net greenhouse effect by clouds, which cannot be captured by our cloud-free 1D model.

The 3D model results by \cite{Shields2014}, for their last habitable warm start scenarios before glaciation, give very low surface temperatures as low as \unit[235]{K}. These surface temperatures are found using the 1D model when assuming a surface albedo between 0.4 and 0.6 (see Fig.~\ref{fig:ToverS}). We find from our comparison of the 1D and 3D model results that a surface albedo of about 0.6 could be considered the boundary between habitable and uninhabitable surface conditions for planets with a large water covered surface area like the Earth, since all habitable scenarios except those by \cite{Yang2014} for slowly rotating planets can be fit with surface albedos smaller than 0.6, and nearly all glaciated scenarios considered here can be approximated with a surface albedo close to 0.6. 

We have shown that for intermediate surface temperatures (between about \unit[250-310]{K}) in the 3D model calculations the 1D model  best approximates these temperatures when assuming a surface albedo of 0.22 and an Earth-like relative humidity profile. This is true even for the coldest habitable scenario from \cite{Kunze2014}, which would usually not be considered as habitable because of their low global mean surface temperature. This is a comforting result as many climate chemistry calculations have been carried out using this combination of relative humidity and surface albedo.

Other sets of surface
albedo and relative humidity are required in the 1D model calculations for scenarios beyond this temperature range, as in the scenarios at the inner edge of the habitable zone by  \cite{Wolf2015} and \cite{Wolf2013}, the warm start scenarios by \cite{Shields2014} close to glaciation, and for the Earth-like planet around the K-type star by \cite{Godolt2015}.

We find that the assumption of a surface albedo of
0.6 in the 1D model best matches the surface temperatures of most glaciated
3D model scenarios for the uninhabitable glaciated scenarios in \cite{Boschi2013}, \cite{Kunze2014}, and \cite{Godolt2015}. This is the case despite the fact that the surface albedo calculated in the 3D model could be higher, such as~for the Ax2.5 scenarios in \cite{Kunze2014}. From this finding one could argue that if surface temperatures calculated with a 1D model with a surface albedo of 0.6 are higher than about \unit[250]{K} then bistable glaciated and habitable climate states are not possible for Earth-like planets with Earth-like compositions.\\

The approximations of the 3D model results by the 1D model calculations with different albedos and relative humidities, however, only hold for the 3D model scenarios that have Earth-like planetary rotation periods. The negative cloud feedback for slowly rotating planets and the resulting surface temperatures (about \unit[310]{K}) at high stellar insolations, as found by \cite{Yang2014}, cannot be approximated with the cloud-free 1D model used here, which calculates a global mean climate and an even heat redistribution between the day and night side. Such temperatures cannot be approximated with a high surface albedo of 0.8 either because the surface albedo is partly masked by the atmosphere. Also, further increasing the surface albedo in the 1D model used here does not lead to a good approximation of the surface temperatures obtained in their 3D modeling study either. Nevertheless, assuming a high surface albedo in the 1D model gives a much wider orbital range at which habitable conditions on the surface 
could be met. This better approximates the 3D model results by \cite{Yang2014} for the slowly rotating planets with negative cloud feedback, than any other assumption of surface albedo in the 1D model. This negative cloud feedback for slowly rotating planets, however, could not have been identified by a 1D model. Describing such new climate phenomena needs self-consistent treatment of the water vapor and cloud feedbacks. Hence, the postulation of a larger width of the HZ estimated by a 1D cloud-free climate model due to the assumption of a surface albedo as high as 0.8 would most probably have been regarded as unrealistic. A surface albedo of 0.8 is representative of a frozen water surface and thus improbable at temperatures above \unit[273]{K}.

\subsection{Relative humidity}

For the relative humidity and, hence, the vertical water vapor distribution, we assumed profiles between an Earth-like relative humidity profile by \cite{MW1967} and a fully saturated atmosphere. From Fig.~\ref{fig:ToverS} we can draw the conclusion that for Earth-like planets around the Sun, as modeled by \cite{Wolf2015} and \cite{Leconte2013}, we could approximate the surface temperatures of the 3D model by assuming a relative humidity in between an Earth-like relative humidity and a fully saturated atmosphere and by assuming a certain surface albedo. We  tested this for the Earth-like planet around the K-type star, as computed by \cite{Godolt2015}, and found that the surface temperatures resulting from the 3D model could also be approximated by assuming a constant relative humidity of 0.8 and a surface albedo between 0.22 and 0.1. The high surface temperatures obtained in their study, however, disagree with the parametrization derived by \cite{Yang2014} 
for rapidly rotating planets. From this parametrization it would be anticipated that an Earth-like planet around a K-type star would be cooler at this stellar insolation. The \cite{Yang2014} inner HZ limit, which corresponds to a surface temperature of about \unit[310]{K}, where their model becomes numerically unstable, is derived to be at \unit[1.14]{S$_{Sun}$} for rapid rotation. Hence future 3D model studies are required to evaluate the temperature response of Earth-like planets around K-type stars. Recently, \cite{delGenio2016} discussed differences in 3D model results at the inner edge of the HZ and argued that especially the parametrization of the convection may lead to different water vapor distributions, and therefore also surface temperatures. \\ 

From the surface temperature increase with stellar insolation found by \cite{Wolf2015} it could be argued that a water vapor profile parametrization that follows the behavior of the saturation pressure curve would be more appropriate than a fixed relative humidity profile similar to \cite{MW1967}. We also tested the relative humidity parameterizations by \cite{Kasting1986} and \cite{Cess1976}, which lead to a change in the relative humidity profile with surface temperature. We found that these parameterizations cannot approximate the surface temperatures of the warm scenarios found by \cite{Wolf2015} and \cite{Godolt2015}. We furthermore constructed relative humidity profiles from the 3D model calculations by \cite{Godolt2015} and found that these lead to a good approximation of the global mean surface temperatures, especially when applied in combination with the surface albedos obtained from the 3D model calculations. More 3D model calculations of different planetary climates are necessary, however, to 
construct a good parametrization of relative humidity for use in 1D models. Preferably, these calculations will come from  different 3D models over a wide range of parameters, such as atmospheric composition, atmospheric mass, planetary gravity, stellar spectra, continental setup, obliquity, and oceanic heat transport, which are beyond the scope of this paper.\\

Despite the fact that most surface temperatures resulting from 3D model calculations can obviously be approximated by 1D model calculations with the appropriate sets of relative humidity and surface albedo, it is also important to investigate whether the global mean temperature profiles can be approximated as well. This was the case for the model calculations in \cite{Godolt2015} . \cite{Wolf2015}, for example, found a temperature inversion in the lower atmospheric layer that could not be captured by the 1D model used here since the convection parametrization does not allow for such an inversion layer. However, a 1D model, as used by \cite{Wordsworth2013}, with another convection scheme could probably reproduce the temperature-pressure profiles of the 3D model study by \cite{Wolf2015}. 
While it is possible to approximate the mean surface temperature of a planet with a well-chosen combination of relative humidity and surface albedo, hence estimating the potential habitability of the planet, other variables may not be comparable, such as the outgoing longwave radiation (OLR), which is a variable that is more easily accessible to observations than the surface temperature of an extrasolar planet. This is caused by the fact that clouds can strongly alter the OLR and the reflected light of an Earth-like planet, as shown, for example, by \cite{Kitzmann2011,Kitzmann2011b}.

\section{Summary and conclusion}
We have investigated the influence of different relative humidities and surface albedos upon the surface temperatures of Earth-like planets as modeled with a 1D cloud-free climate model. Relative humidity and surface albedo are the result of climate feedbacks, which are usually not included in 1D models. Nevertheless, since relative humidity and surface albedo can in principle vary on an Earth-like planet only over a limited range, 1D parameter studies could in principle estimate the range of climates that an Earth-like planet may have at a certain stellar insolation. \\

We compared the result of our 1D model calculations to those of 3D model calculations in the literature. We showed that for most of the 3D model calculations a combination of relative humidity and surface albedo can be found to approximate the resulting surface temperatures with a 1D model. This also shows that the larger extent of the habitable zone, as suggested by the studies of \cite{Leconte2013} and \cite{Wolf2015}, can also be found by 1D models. The expansion of the habitable zone, as suggested by \cite{Yang2014}, via a negative cloud feedback for slowly rotating planets can be partly approximated by the cloud-free 1D model by assuming very high surface albedos. However, the surface temperatures obtained with the 1D model deviate from those of the 3D model and most probably the mean vertical temperature structures as well. Therefore, the approach of mimicking high cloud refection by increased surface albedo should be handled with care.    \\ 

For intermediate surface temperatures of habitable scenarios ranging from about 250 to \unit[310]{K} in the 3D model calculations, the 1D model used here gives the best approximations when assuming an Earth-like relative humidity \citep{MW1967} and surface albedo (0.22) that reproduce the surface temperature of the present Earth. For higher surface temperatures, the relative humidity in the 1D model  has to be increased\ to match the global mean surface temperatures of the 3D model calculations. Surface temperatures of uninhabitable glaciated 3D model scenarios, which can occur at the same stellar insolations as habitable scenarios due to different boundary or initial conditions, are best approximated when assuming a mean surface albedo of 0.6 in the 1D model used here. At these low temperatures, the relative humidity chosen does not show a large impact on the surface temperatures.\\

Future comparisons of 1D and 3D climate calculations may allow the construction of relative humidity parameterizations for use in 1D models to better approximate the water vapor feedback at higher surface temperatures. However, more 3D model calculations are needed covering additional scenarios with different models over extended boundary and initial conditions. 
Until such parameterizations are at hand and to explore the possibility of bistable climates, we suggest varying relative humidity and surface albedo when, for example, evaluating the habitability of a planet or calculating the photochemical and spectral response of Earth-like planets around different types of stars. Furthermore, the temperature range that is considered as potentially habitable needs to be extended down to much lower temperatures.

\begin{acknowledgements}
      Part of this work was supported by the Helmholtz Gemeinschaft, project
      number PD-015.
\end{acknowledgements}

%
  \bibliographystyle{aa} 
  \bibliography{RH_Albedo} 
%

\end{document}